\documentclass[10pt,a4paper,twoside]{article}
\usepackage[utf8]{inputenc}
\usepackage{graphicx}
\usepackage{nicefrac}
\usepackage{booktabs}
\usepackage{float}
\usepackage{color}
\usepackage[left=2cm,right=2cm,top=2cm,bottom=2cm]{geometry}

\author{Paulina Trybek$^1*$, Michal Nowakowski$^2$, Lukasz Machura$^{1}$, }
\date{}
\title{Multifractal characteristics of external anal sphincter based on sEMG signals}

\begin{document}
\maketitle
\noindent $^1$ Division of Computational Physics and Electronics, Institute of Physics, Silesian Center for Education and Interdisciplinary Research, University of Silesia, Chorzow, Poland
\\
$^2$ Department of Medical Education, Jagiellonian University Medical College, Krakow, Poland
\\
$^*$ paulina.trybek@smcebi.edu.pl

\section*{Abstract}

This work presents the application of 
Multifractal Detrended Fluctuation Analysis for the surface electromyography 
signals obtained from the patients suffering from rectal cancer. The 
electrical activity of an external anal sphincter at different levels of medical
treatment is considered.
The results from standard MFDFA and the EMD--based MFDFA method are compared.
Two distinct scaling regions were identified. Within the region of short time scales
the calculated spectra exhibit the shift towards higher values of the
singularity exponent for both methods. In addition obtained spectra are 
shifted towards the lower values of singularity exponent for the 
EMD--based MFDFA.

\section{Introduction}
\label{intro}
Over last few decades a surface electromyography due to its non--invasive character has gained a wide range of applications in the field of  examination of a neuromuscular system.
This work is focused on quite a unique  application of the sEMG concerning a diagnosis of anal sphincter of the patients suffering from a rectal cancer \cite{nowakowski2014developing,merletti2004multichannel}. Rectal cancer remains to be one of the most frequent cancers in humans \cite{bray2012global}. 
It requires complex multimodal treatment composed of surgery, irradiation 
and chemotherapy. All those methods can cause significant 
stool continence related problems hence proper assessment of anorecatal 
innervation before and after the treatment can be crucial for prevention 
and treatment of complications. sEMG enables non-invasive monitoring of the anal sphincter function 
\cite{enck2005external,Kauff,Cescon} and is a very promising method of testing of innervation of muscles. Since 
innervation deficits are one of suggested mechanisms for severe treatment related complications in up to 40\% of 
rectal cancer patients development of proper innervation assessment methodology is crucial. 
Proper evaluation of sEMG signals remains to be a significant 
problem inhibiting diagnostic potential of this methodology. Our analysis is focused on the signals recorded from the 
external anal sphincter. The electrical activity of this specific muscle group is frequently investigated in the 
context of the patients with defecation disorders \cite{lopez1999electromyography}. 

Regardless the application, sEMG always 
represents highly complex signal with a low signal to noise ratio \cite{clancy2002sampling}. The nonlinearity of sEMG 
data has been investigated in recent years \cite{lei2001detecting} and great effort has been devoted to the 
application the variety of nonlinear  methods. Traditional analysis, mainly based on the conventional statistical 
tests of mean, median or frequency components brings only limited knowledge on the actual process hidden
behind the acquired data. 

In the last few years there has been a growing interest in fractal properties of physiological data, also in the 
context of sEMG signals  \cite{goldberger2002fractal, atupelage2012multifractal, wang2007multifractal}.  This work 
propose the application of the modified Detrended Fluctuation Analysis (MFDFA) based on Empirical Mode Decomposition 
(EMD) to the sEMG signals. The EMD and MFDFA  techniques can be 
used to trace out the features of non--linear and non--stationary signals. Moreover
both methods have a broad spectrum of applications individually. 
MFDFA, developed by Kantelhardt et al. \cite{kantelhardt2002multifractal} has been used in many disciplines and still 
attracts considerable attention in the field of physiology, economy, climatology, to name 
but a few. In relation to the electrophysiological signals, MFDFA brought a significant contribution  to the analysis 
of heart rate variability \cite{makowiec2011reading,gieraltowski2012multiscale}. 
For the Empirical Mode Decomposition (EMD) the equally wide range of applications can be found, such as the removal of 
artifacts and noise reduction from the signals \cite{andrade2006emg}. EMD also exhibits better results in the process 
of detrending  in comparison for example with the typically used least square method \cite{liu2015precipitation}. This 
aspect has been used in the modified detrending algorithm which is presented in this paper. The use of the EMD method 
in the context of detrending operation results in a more accurate trend, which is not predetermined, and therefore is 
closely related to the nature of real data \cite{yeh2009human}. Moreover 
it is documented in literature that this approach outperforms standard MFDFA for large fluctuations 
\cite{qian2011modified}.

The paper is organized as follows. Section \ref{s1} presents both detrending methods and a multifractal formalism
for data analysis. Next section \ref{s2} describes the experimental data. The results are presented in 
Section \ref{s3}. The last section summarizes the results and draws  conclusions.

\section{Method}\label{s1}

\subsection{DFA}\label{ss11}
DFA method was first proposed by Peng in 1994 for investigating the correlation in DNA structure 
\cite{peng1994mosaic}. The last years have seen a renewed importance in application of the method to the biological 
data, also for distinguish  healthy 
and pathological states \cite{rodriguez2007detrended}.  The basic idea of this technique relies on the assumption that 
signal is influenced by the short--term and long--term features. For the proper interpretation of effects 
hidden behind 
internal dynamics the signal is analyzed at multiple scales \cite{semmlow2014biosignal}.  
The brief description of the original DFA algorithm is presented below.

The procedure starts with calculation of the profile $y_i$ as the cumulative sum 
of the data $x_i$ with the subtracted mean $\langle x \rangle$:
\begin{equation}
 y_i = \sum_{k=1}^i [x_k - \langle x \rangle]   
\end{equation}

Next, the cumulative signal $y_i$ is split into $N_s$ equal non-overlapping segments of
size $s$. Here for the length $s$ of the segments we use powers of two, $s = 2^r, r=4\dots11$.
For all segments $v = 1, \ldots, N_s$ the local trend $y^m_{v,i}$ is calculated. 
In a standard DFA method the trend is calculated by means of the least--square fit of order $m$.
In this work $m=2$ was used. The variance $F^2$ as a function of the segment length $s$
is calculated for each segment $v$ separately. 
\begin{equation} 
F^2(s,v) \equiv \frac{1}{s} \sum_{i=1}^{s} 
\left( y^m_{v,i} - y_{v,i} \right)^2. \label{fsdef} 
\end{equation}

For the last step the Hurst scaling exponent $H$ is calculated as the slope of the regression line 
of double-logarithmic dependence, $\log F \sim H \log s$.

\subsection{Empirical Mode Decomposition (EMD)}\label{ss12}

The EMD is an iterative technique which decomposes the signal x(t) into finite number of 
Intrinsig Mode Functions (IMFs) $c_i(t)$. The final 
residual $r_n(t)$ can be interpreted as an actual trend. 
\begin{equation}
x(t) = \sum_{i=1}^{n}c_i(t)+r_n(t)
\label{imfs}
\end{equation}
The calculated signal must satisfy two conditions in order to be IMF: 
(i) the number of extrema and the number of zero crossing must
be equal or differ at most by one; and (ii) the mean value of
the upper and lower envelope defined by local maxima and minima must be zero. 
The standard EMD method often faces some difficulties, which are recurrently 
the consequence of signal intermittency referred to as 
Mode Mixing problem. Ensemble Empirical Mode Decomposition (EEMD) \cite{wu2009ensemble}  and more recent Complete Ensemble Empirical Mode Decomposition (CEEMD) \cite{torres2011complete} 
 have been proposed in order to overcome this complication. 
Both methods are based on the averaging over several realisations of 
Gaussian white noise artificially added to the original signal. 
For this work, however, we use only standard EMD due to the fact that only 
residual $r_n$, i.e. the data trend, is needed for further calculations 
and none of the individual IMFs are considered here explicitly.

\subsection{EMD based DFA}\label{ss13}
\begin{figure}[htbp]
    \centering
    \includegraphics{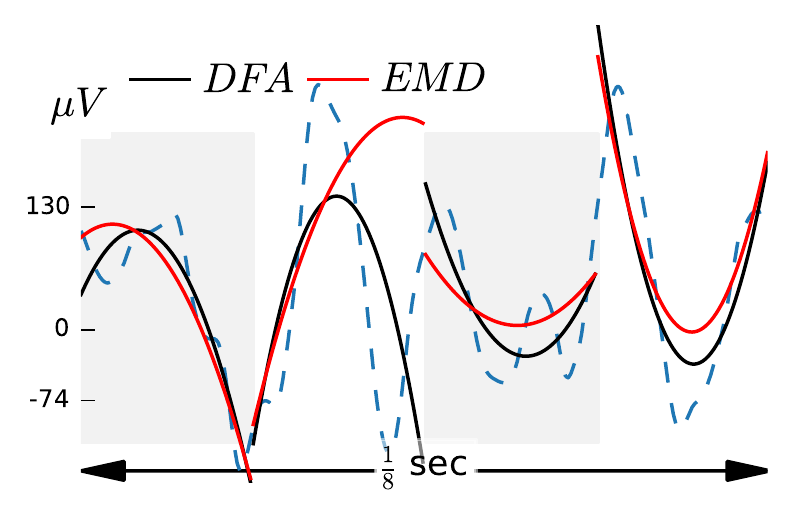}
    \caption{Two detrending methods: DFA (solid black) and EMD (solid red) 
    are presented 
    for the profile $y_i$ of the sEMG example data (dashed blue).}
    \label{fig1}
\end{figure}

Our analysis was branched into standard DFA algorithm and non-standard one based on EMD technique. 
The former method uses the least--square estimation of the order $m$. The latter utilizes the fact 
that the residual $r_n$ (\ref{imfs}) represents the local trend, thus the standard polynomial 
fit (DFA) can be replaced by a residuum for each segment
\cite{caraiani2012evidence}. An example of local trends 
calculated with both methods is presented on Fig.\ref{fig1} for the segment size 
$s=64$. The slight differences between solid black and red lines, which represent 
DFA and EMD method, respectively, influence the further results.

\subsection{MFDFA}\label{ss14}
MFDFA is based on the scaling properties of the fluctuations. 
The brief description of the method is presented below, however, for 
the detailed specification we suggest works by Kantelhardt et al. 
\cite{kantelhardt2002multifractal, kantelhardt2012fractal}, Ihlen \cite{ihlen2012introduction} or Salat et al. \cite{salat2017multifractal}.
In order to extend the monofractal DFA (\ref{ss11}) to the multifractal DFA it is necesarry to indicate  the $q^{th}$ statistical moment of the calculated variance (\ref{fsdef}).
The equations that describe the fluctuation functions are presented below.

\begin{equation}\label{fluf}
F_q(s) = \left\{
  \begin{array}{lr}
    \left(\frac{1}{2N_s}\sum_{v=1}^{2N_s} [F^{2}(s,v)]^\frac{q}{2} \right)^\frac{1}{q}, &  q \ne 0,\\
    \exp \left\{ \frac{1}{4 N_s}\sum_{v=1}^{2 N_s} \ln \left[F^2(s,v)\right] \right\}, & q = 0.
  \end{array}
\right.
\end{equation}

Next, the determination of the scaling law $F_q(s) \sim s^{h(q)}$ of the fluctuation 
function (\ref{fluf}) is performed with the use of the log--log dependencies of $F_q(s)$ versus segment 
sizes $s$ for all values of $q$ separately.
The $q$-order Hurst exponent $h(q)$ is required in order to calculate the further dependencies. 
The mass exponent is obtained via the formula
\begin{equation}
\tau(q) = q h(q) - 1. 
\end{equation}
It is than used to calculate a $q$--order 
singularity H\"older exponent $\alpha = \tau'(q)$ where the prime
means differentiation with respect to the argument. 
In turn the $q$--order singularity dimension can be constructed
\begin{equation}\label{spectrum}
f(\alpha) = q \alpha - \tau(q) = q[\alpha - h(q)] + 1.
\end{equation}
The singularity dimension $f(\alpha)$ is related to the mass 
exponent $\tau(q)$ by a Legendre transform. The multifractal spectrum,
i.e. the dependence $f(\alpha)$ vs $\alpha$ is the final result of MFDFA method.

\section{Material}\label{s2}
\subsection{sEMG signal source}\label{ss21}

Data acquisition system consists of anal probe developed at Laboratory of Engineering of Neuromuscular System and Motor Rehabilitation of Politecnico di Torino in collaboration with the company OT-Bioelettronica.
Signals were obtained from 16 pairs of silver bar electrodes of length 9 mm and width 1
mm each. Electrodes were separated by 8 mm and arranged concentrically at three 
levels 35--44 mm, 18--27 mm and around 9 mm of rectal canal depth from the anal verge. 
The probe worked in conjunction with the standard PC over 12 bit NI DAQ MIO16 E-10 transducer (National Instruments, USA). 
The sampling frequency was 2048 Hz, which for the 10 seconds of the measurement gave 20480 data points.  
Low and high pass filters  were used at 10 and 50 Hz respectively. This resulted in typical 3dB bandwidth for the ADC.
The analyzed time series were recorded at four stages: before the treatment ($D_1$) and one month ($D_2$), six months 
($D_3$) and one year ($D_4$) after the surgical procedure. The detailed information about the surgery of the rectal 
cancer and the role of sEMG for the patient diagnosis can be found in \cite{delaini2005functional}.

\begin{figure}[htbp]
\centering
\includegraphics[width=0.45\linewidth]{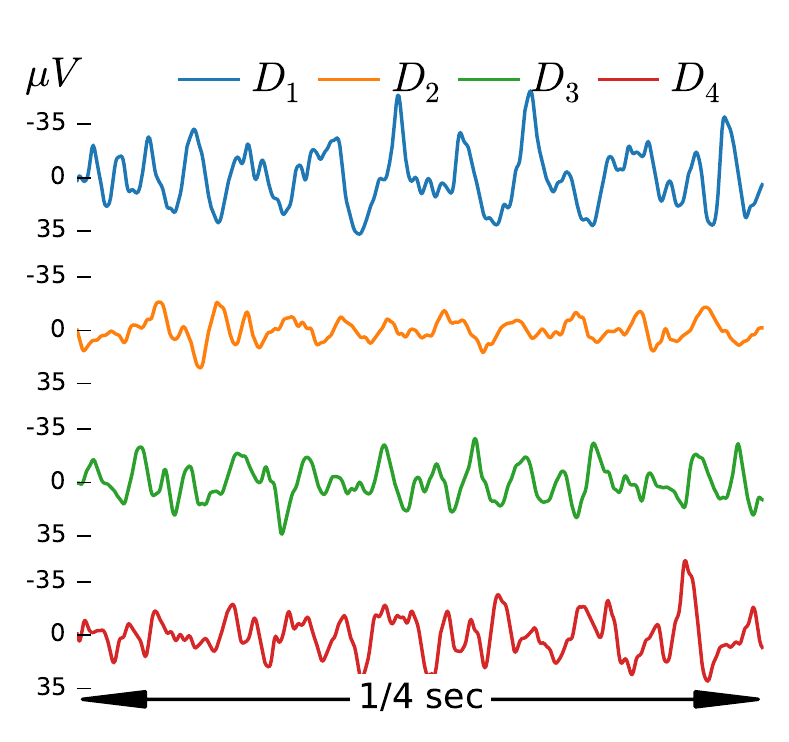}
 \caption{The raw signals truncated to \nicefrac{1}{4} of the second 
    at four stages of rectal cancer treatment: 
    $D_1$ is assigned to the state before surgery, $D_2$ -- $D_4$ correspond 
    to one month, six months and one year after the surgery, respectively.}
\label{fig2}
\end{figure}

Measurement protocol included consecutively 1 minute 
relaxation state, three 10 sec recordings at rest and 1 minute relaxation and 
then three 10 sec recordings at maximum voluntary contraction (MVC). Each signal was recorded at three levels of 
anal canal depth, respectively 5cm, 3cm and 1cm.  For our calculations signal recorded during voluntary contraction at 
the depth of 1 cm was used. This specific choice of depth was dictated by the maximal amplitude of 
EMG signal resulting 
from the most superficial localisation of external anal sphincter muscle and the biggest bulk of the muscle at this 
location. Fig. \ref{fig2} presents an example of the raw signals at four stages of rectal cancer treatment. For better 
visualization of the character of waveforms, 
the narrow range of time scale is presented. The difference in the amplitude 
values between the state directly after operation ($D_2$) and the rest of 
states ($D_1$, $D_3$, $D_4$) is visible to the 
naked eye. One month after the operation the values of amplitude are respectively 
lower in comparison with other stages. 
The signal used in this 
work is in fact the averaged signal from the first 3 channels 
which corresponds to the first three pairs of the electrodes. The nearest 
neighbours average was performed due to the fact that the placement of the 
probe in subsequent measurements could be inaccurate. In other words the 
specific electrodes may not be located exactly the same place at the 
consecutive measurements after the surgery.

\subsection{Patients}\label{ss22}
The study included 15 subjects, 5 female, age range 46 to 71 (average 53,4 years) and 10
male, age range 40 to 85 (average 62,8 years), diagnosed with rectal cancer and qualified
for surgery. Based on localization of rectal cancer patients underwent either Low
Anterior Resection (LAR, 9 patients), Anterior Resection (AR, 5 patients) or
proctocolectomy (PC, 1 patient).

\section{Results}\label{s3}
\subsection{Fluctuation Analysis}\label{ss31}
\begin{figure}[htbp]
\centering
\includegraphics{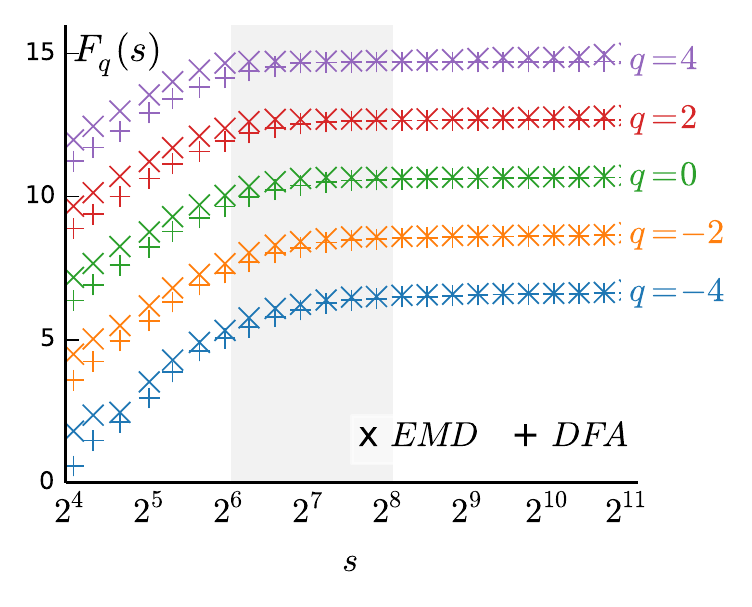}
\caption{$q$-th order fluctuation function (\ref{fluf})
with both detrending methods DFA of order 2 and EMD
presented for data before the surgery $D_1$
for selected values of $q$. Characteristics were artificially 
shifted vertically for better visibility.}
\label{fig3}
\end{figure}

Clearly, MFDFA is not a black--box method and always requires some individual 
decisions. First of all, the choice of the scaling range can have a significant 
impact on the appropriate estimation of the fluctuation function ($F_q$) and 
consequently the final results \cite{gieraltowski2012multiscale, ihlen2012introduction}. 
For the calculations presented in this work, the considered range 
of scales are between $s \in [2^4,2^{11}]$. The parameter $q$ should consist of 
positive and negative values in order to detect periods with 
small and large fluctuations \cite{ihlen2012introduction}. In our case 
$q \in [-5, 5]$ were chosen. A set of $q$-order fluctuation functions $F_q$ 
\textit{vs} segment size $s$ is presented in Fig.\ref{fig3}. The two 
different scale ranges are clearly visible for all $F_q(s)$ characteristics. 
This bisection into two distinct scaling regimes plays a crucial role in determination of the
$q$-order Hurst exponent $h(q)$ and wherefore impacts the further analysis.
The results for DFA and EMD--based detrending method are presented on Fig. \ref{fig3}. 
Two separate scaling domains was accepted, namely $s \in [2^4, 2^6]$ and 
$s \in [2^8, 2^{11}]$. Further analysis have been performed for both
of this regions separately. The middle values $s \in (2^6, 2^8)$ are omitted, 
as there is no clear linear scaling present.

\subsection{Multifractal spectra}\label{ss32}
\begin{figure}[h]
    \centering
    \includegraphics[width=0.9\linewidth]{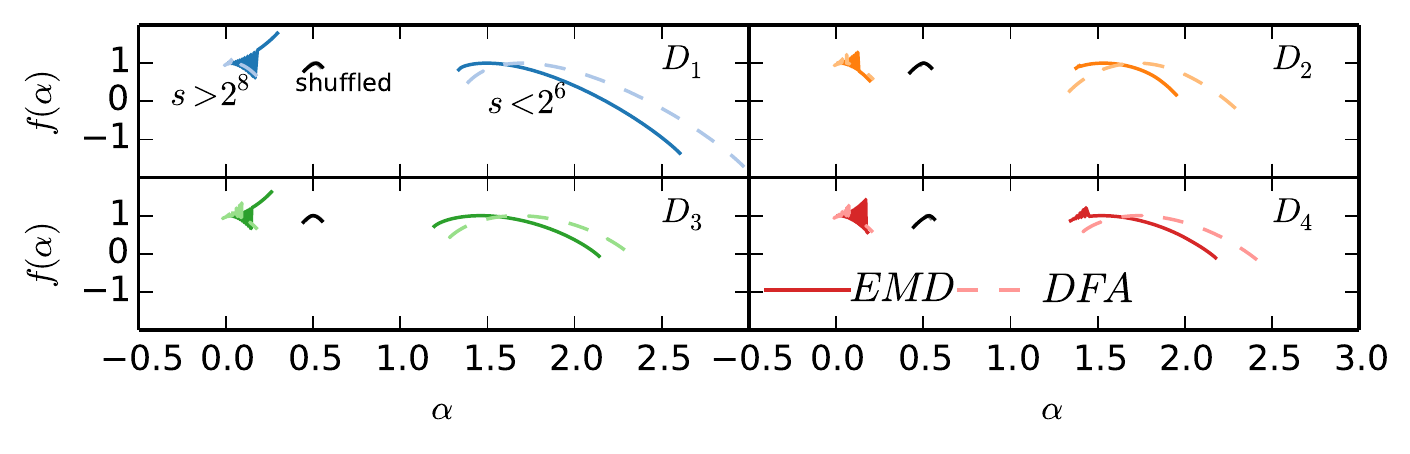}
    \caption{An example of multifractal spectra calculated for one case at four stages of 
    rectal cancer treatment $D_1$ -- $D_4$. At each panel three sets of 
    distinct spectra can be found: right which correspond to the scaling
    region $s < 2^6$, the middle represent the whole range of scales $s \in [2^4,2^{11}]$ after shuffling operation 
    and the left set for $s>2^8$. For each scaling regions
    two spectra are presented -- for the EMD based MFDFA (dark solid lines)
    and standard MFDFA (light dashed lines). In the case of shuffled data the spectra calculated by both methods are overlapped almost entirely. One can notice the generally
    found degeneracy of the spectra for $s > 2^8$ and the shift towards
    the smaller values of $\alpha$ for the EMD--based detrending.
    }
    \label{fig4}

\end{figure}

The central result of this work is presented in Fig.\ref{fig4}. 
It sets together the multifractal spectra at all level of treatment process ($D_1$-$D_4$).
The mf--spectrum describes how often the 
irregularity of certain degree occurs in the signal. $f(\alpha)$ represent $q$-order singularity dimension and 
$\alpha$ stands for the $q$-order singularity exponent.  The typical monofractal time series has dense mf--spectrum 
around the single point $(\alpha=0,f(\alpha)=1)$. A large difference between periods when small and large fluctuations 
takes place increases in turn the width of the spectrum.

There are two general sources of multifractality which can affect the shape of the mf-spectrum:
(i) broad probability density function which lies behind the data (or its
fluctuations); (ii) different behaviour of the (auto)correlation function for
large and small fluctuations; (iii) both situations simultaneously.
Simple data shuffling can test the possible source of multifractality. In the case (i)
shuffling will not change the mf-spectrum, for (ii) will destroy the effect completely
as the shuffling will erase the possible correlations. In the last case (iii) the spectrum
will differ from the original one as the shuffled series will exhibit somehow weaker multifractality 
\cite{horvatic2011detrended}. 
The spectra calculated for shuffled data (see black line in  Fig. \ref{fig4}) form a tight set of points, 
significantly shifted towards the lower values of singularity exponent $\alpha$, in other words the 
shuffling operation resulted in a complete destruction of multifractality which occurs for the raw data. 
Thus it can be concluded that multifractal character of raw data has its cause in different behaviour of 
the correlation function for large and small fluctuations.


For all of the examined cases the relatively wide spectra for the short scales $s<2^6$ can be observed. 
For the large scales $s>2^8$ the small set of points located around point $(0, 1)$ is visible, in that case both multifractal methods result in degenerated narrow spectra for all stages of treatment. This indicates the multifractal character of the sEMG signal for the short
scales $s<2^6$ and rather monofractal character for the large scales $s> 2^8$.  For all of the presented analysis, regardless the method, the spectra calculated for 
small scaling region $s < 2^6$, exhibit
long right tails, see Fig. \ref{fig4}. It means that the multifractal structure is 
sensitive to the local fluctuations with small magnitudes on the short time scales 
only \cite{ihlen2012introduction}.
On the comparison of the spectra obtained by the two methods a shift towards 
the smaller values of $\alpha$ (left side) of the spectrum for the small scales $s < 2^6$
is visible for all signals in the case of the EMD--based MFDFA.

\subsection{Statistics of spectral parameters}

Tab. \ref{tab:width} summarizes the average values of the spectrum width $\langle \Delta \rangle$  
and the specific singularity exponent $f(\alpha_{max}) = 1$.
which corresponds to the maximum of the spectrum calculated for all of the analyzed cases and the whole set of 16 electrodes at each treatment state.

\begin{table}[h]
\centering
\caption{Average values of the spectrum width 
$\langle \Delta \rangle$ and maximum of spectrum $\alpha_{max}$ 
together with the standard deviations presented 
for all channels at each state of the treatment $D_1$ -- $D_4$. 
Results are presented for MFDFA and EMD--based MFDFA method.}
\label{tab:width}
\begin{tabular}{|c |c | c | c | c|} 
\hline
\multicolumn{5}{|c|}{Average spectrum width $\langle \Delta \rangle$ for  s $< 2^6$ }   \\      \hline
 & $D_1$ & $D_2$ & $D_3$ & $D_4$ \\
\hline
DFA &  \begin{tabular}[c]{@{}c@{}}  $0.981$\\ $\pm$ $0.362$ \end{tabular} &  \begin{tabular}[c]{@{}c@{}}  $0.950$\\ $\pm$ $0.342$ \end{tabular} &  \begin{tabular}[c]{@{}c@{}}  $1.053$\\ $\pm$ $0.418$ \end{tabular}  & \begin{tabular}[c]{@{}c@{}}  $0.967$\\ $\pm$ $0.355$ \end{tabular} \\
\hline
EMD & \begin{tabular}[c]{@{}c@{}}  $0.848$\\ $\pm$ $0.316$ \end{tabular} & \begin{tabular}[c]{@{}c@{}}  $0.808$\\ $\pm$ $0.310$ \end{tabular} & \begin{tabular}[c]{@{}c@{}}  $0.880$\\ $\pm$ $0.350$ \end{tabular} & \begin{tabular}[c]{@{}c@{}}  $0.812$\\ $\pm$ $0.307$ \end{tabular}   \\
\hline
\end{tabular}
\\
\begin{tabular}{|c |c | c | c | c|} 
\hline
\multicolumn{5}{|c|}{Average spectrum width $\langle \Delta \rangle$ for  s $>2^8$ }   \\      \hline
DFA &  \begin{tabular}[c]{@{}c@{}}$0.136$\\ $\pm$ $0.068$ \end{tabular} &  \begin{tabular}[c]{@{}c@{}}$0.142$\\ $\pm$ $0.052$ \end{tabular} &  \begin{tabular}[c]{@{}c@{}}$0.130$\\ $\pm$ $0.055$ \end{tabular}  & \begin{tabular}[c]{@{}c@{}}$0.161$\\ $\pm$ $0.118$ \end{tabular} \\
\hline
EMD & \begin{tabular}[c]{@{}c@{}}  $0.107$\\ $\pm$ $0.070$ \end{tabular} & \begin{tabular}[c]{@{}c@{}}  $0.112$\\ $\pm$ $0.053$ \end{tabular} & \begin{tabular}[c]{@{}c@{}}  $0.100$\\ $\pm$ $0.054$ \end{tabular} & \begin{tabular}[c]{@{}c@{}}  $0.131$\\ $\pm$ $0.118$ \end{tabular}   \\
\hline
\end{tabular}

\begin{tabular}{|c |c | c | c | c|} 
\hline
\multicolumn{5}{|c|}{Maximum of the spectrum $\alpha_{max}$ for  s $<2^6$ } \\ \hline
DFA &  \begin{tabular}[c]{@{}c@{}}  $1.580$\\ $\pm$ $0.107$ \end{tabular} &  \begin{tabular}[c]{@{}c@{}}  $1.577$\\ $\pm$ $0.117$ \end{tabular} &  \begin{tabular}[c]{@{}c@{}}  $1.588$\\ $\pm$ $0.110$ \end{tabular}  & \begin{tabular}[c]{@{}c@{}}  $1.593$\\ $\pm$ $0.104$ \end{tabular} \\
\hline
EMD & \begin{tabular}[c]{@{}c@{}}  $1.405$\\ $\pm$ $0.098$ \end{tabular} & \begin{tabular}[c]{@{}c@{}}  $1.414$\\ $\pm$ $0.100$ \end{tabular} & \begin{tabular}[c]{@{}c@{}}  $1.414$\\ $\pm$ $0.096$ \end{tabular} &  \begin{tabular}[c]{@{}c@{}}  $1.423$\\ $\pm$ $0.089$ \end{tabular} \\
\hline
\end{tabular}

\begin{tabular}{|c |c | c | c | c|} 
\hline
\multicolumn{5}{|c|}{Maximum of the spectrum $\alpha_{max}$ for  s $>2^8$ }   \\     \hline
DFA &  \begin{tabular}[c]{@{}c@{}}$0.030$\\ $\pm$ $0.011$ \end{tabular} &  \begin{tabular}[c]{@{}c@{}}$0.031$\\ $\pm$ $0.008$ \end{tabular} &  \begin{tabular}[c]{@{}c@{}}$0.031$\\ $\pm$ $0.012$ \end{tabular}  & \begin{tabular}[c]{@{}c@{}}$0.031$\\ $\pm$ $0.009$ \end{tabular} \\
\hline
EMD & \begin{tabular}[c]{@{}c@{}}  $0.017$\\ $\pm$ $0.009$ \end{tabular} & \begin{tabular}[c]{@{}c@{}}  $0.017$\\ $\pm$ $0.007$ \end{tabular} & \begin{tabular}[c]{@{}c@{}}  $0.017$\\ $\pm$ $0.010$ \end{tabular} & \begin{tabular}[c]{@{}c@{}}  $0.017$\\ $\pm$ $0.008$ \end{tabular}   \\
\hline
\end{tabular}
\end{table}

\begin{figure}[h!]
\centering
    \includegraphics[width=0.95\linewidth]{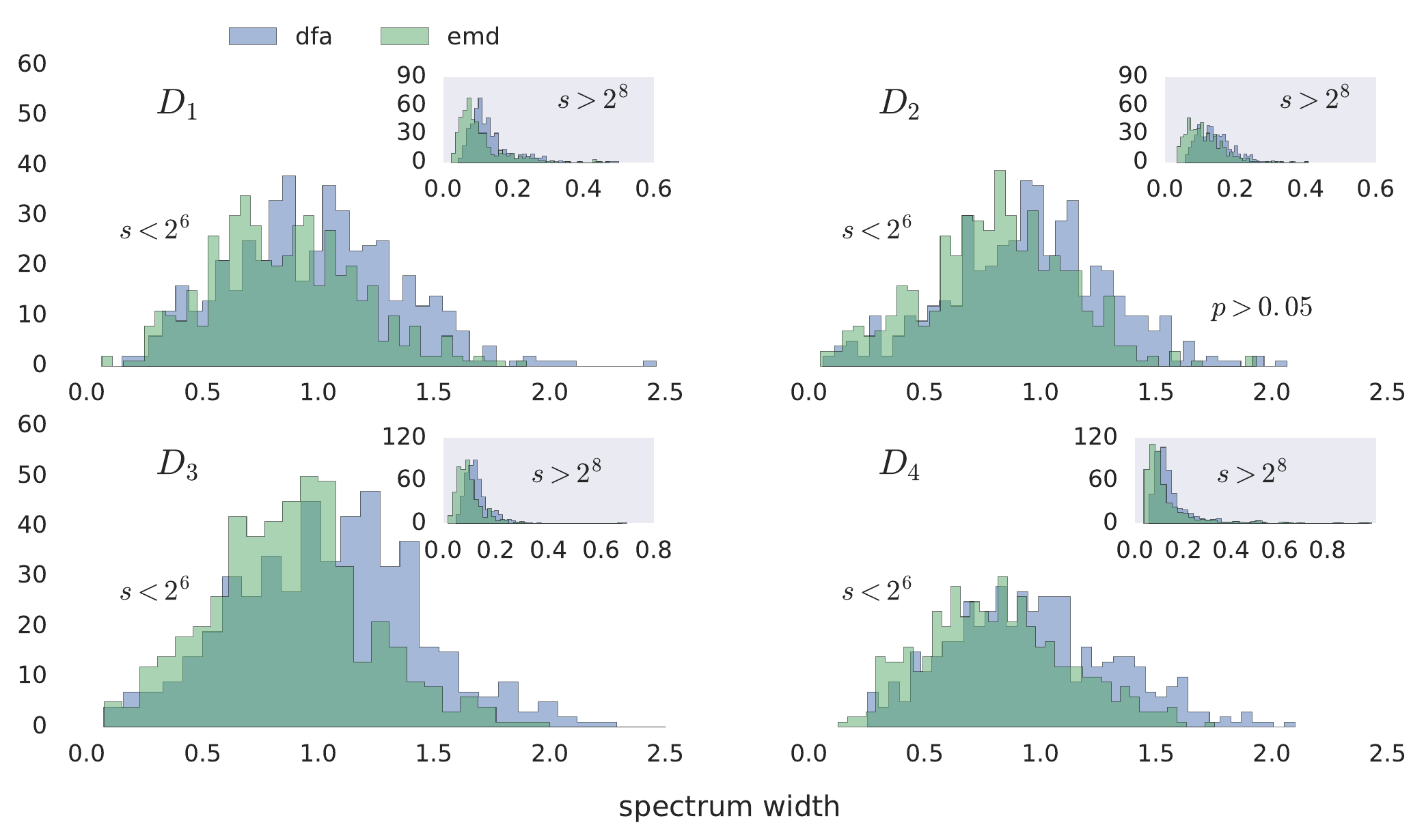}
     \includegraphics[width=0.95\linewidth]{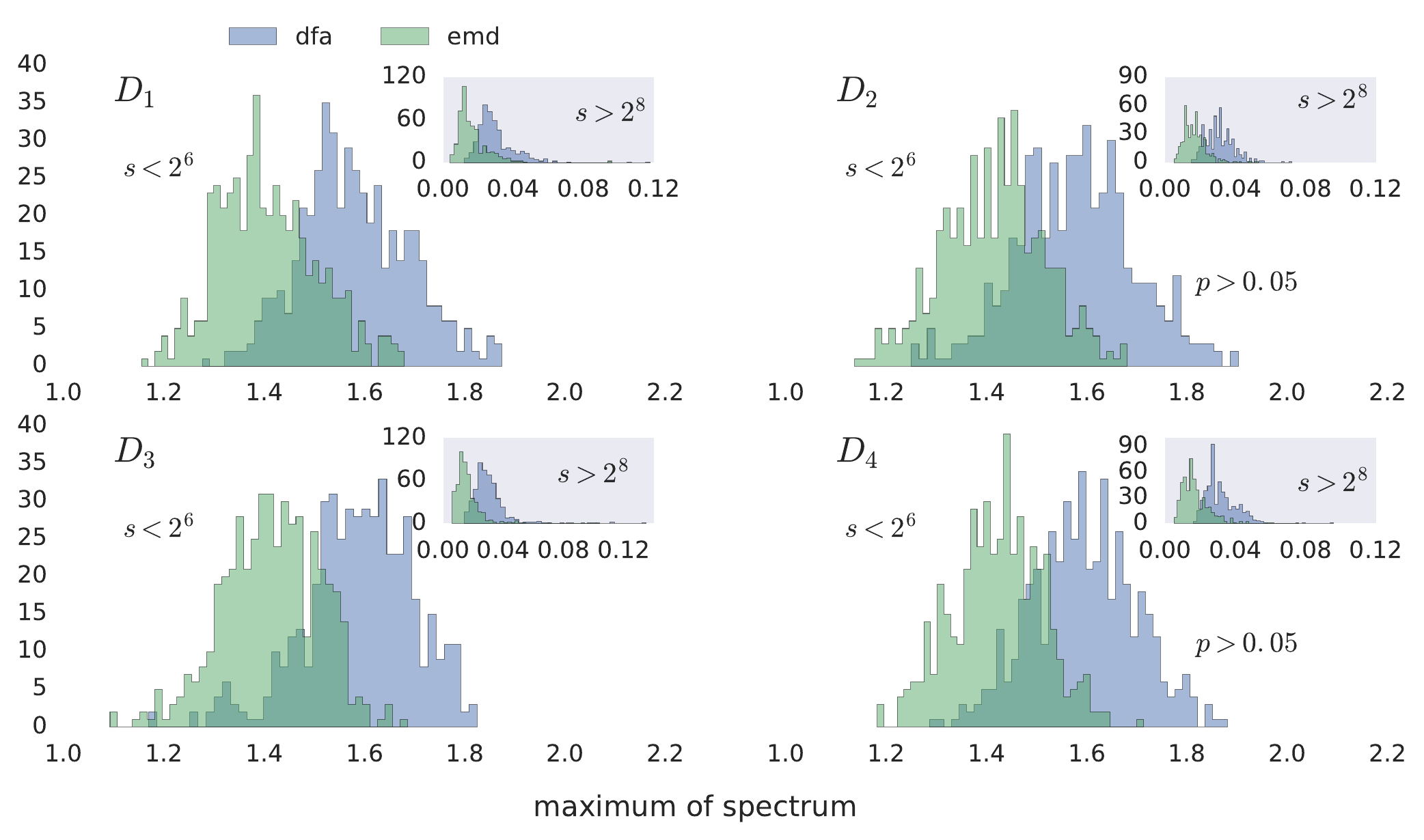}
     \caption{The probability distributions of spectral parameters calculated for both EMD and DFA methods at each state of treatment: large histograms represent the small scaling region $s<2^6$; the insets stand for the histograms corresponding to the 
     large range of scales.     }
    \label{fig5}
\end{figure}

The location of the maximum of the spectrum is always found for the greater 
values of the singularity exponent in the case of standard MFDFA method.
Also the width of the spectrum is consistently wider for the standard MFDFA. 
The shift towards the higher values of spectrum parameters for the standard MFDFA is also visible on the presented histograms, see Fig. \ref{fig5}. The differences are more evident in the graphs that characterize the maximum of the spectrum for both small and large range of scales. 
The normality tests of presented probability distributions by means of the Shapiro--Wilk formula do not allow to reject the hypothesis of normality for some selected cases. At the chosen significance level $\alpha=0.05$, $p$--value is always greater than $\alpha$ for both the spectrum width and the maximum of spectrum for small scaling range ($s<2^6$) at the state one month after surgical procedure ($D2$). Additionally, the same results were obtained for the state one year after the operation ($D4$), however, for maximum of the spectrum only.

\begin{figure}[h]
\includegraphics[width=\linewidth]{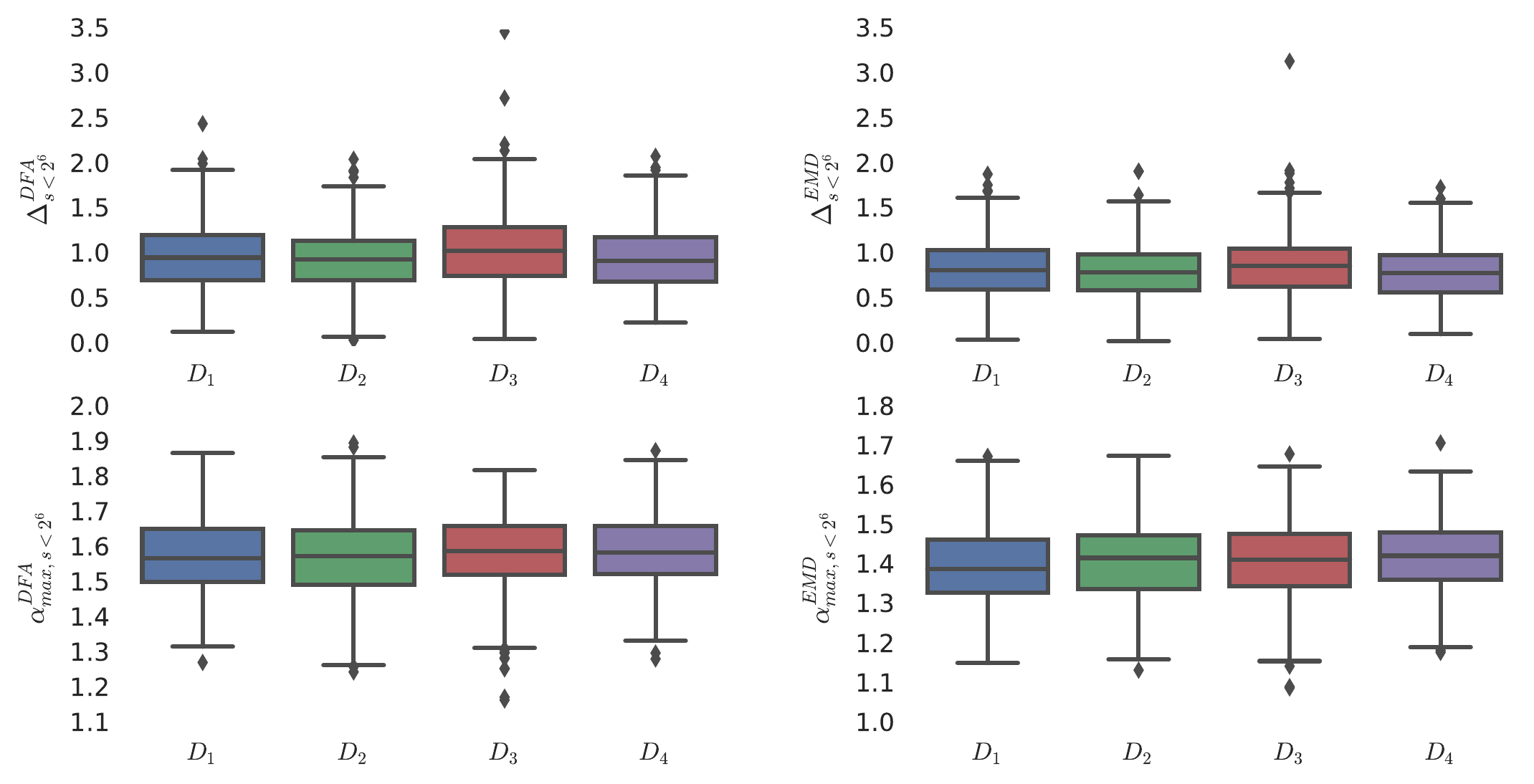}
\caption{The box plots of spectral parameters calculated for the small range of scales. The box shows the quartiles of the detests whereas points are assigned to outliers values. }
\label{fig6}
\end{figure}

Fig. \ref{fig6} presents the box plot of spectral parameters calculated for small scaling region $s<2^6$.
The comparison of mf-spectrum parameters for different states of treatment show a 
slight decrease in the average value of the spectrum maximum $\alpha_{max}$ and 
spectrum width $\langle \Delta \rangle$ for the state one month after the 
surgery -- $D2$. The results of the $p$--value calculated for nonparametric Wilcoxon test are presented in table \ref{table2}. The values highlighted in red characterize the statistical significance of the difference between comparing stages. It is noticeable that the results for both methods are ambiguous. In the case of spectrum width, the consistent results are obtained for the state 6 month after surgery ($D_3$). This state differs from all the others at the selected significance level ($\alpha = 0.05$). Considering the result of $\alpha_{max}$, the statistically significant difference indicated simultaneously by DFA and EMD occurs between states before and one year after the surgery ($D_1-D_4$).

\begin{table*}[h!]\centering
\begin{tabular}{@{}rrrcrrrcrr@{}}\toprule 
& \multicolumn{2}{c}{spectrum width($\Delta$)} & \phantom{abc}& \multicolumn{2}{c}{maximum of spectrum($\alpha_{max}$)} &
\phantom{abc}\\
& DFA & EMD  && DFA & EMD \\ \midrule
 Compared stages\\
\large{$D_1$--$D_2$} & 0.089 & \textcolor{red}{0.014} && 0.881 & \textcolor{red}{0.009} \\
\large{$D_1$--$D_3$} & \textcolor{red}{0.001} & \textcolor{red}{0.027} && 0.294 & 0.337\\
\large{$D_1$--$D_4$} & 0.754 & \textcolor{red}{0.031} && \textcolor{red}{0.006} & \textcolor{red}{0.000} \\
\large{$D_2$--$D_3$} & \textcolor{red}{0.000} & \textcolor{red}{0.000} && \textcolor{red}{0.007} & 0.66 \\
\large{$D_2$--$D_4$} & 0.341 & 0.508 && \textcolor{red}{0.022} & 0.424 \\
\large{$D_3$--$D_4$} & \textcolor{red}{0.000}& \textcolor{red}{0.000}  && 0.215 & \textcolor{red}{0.012}\\
\bottomrule
\end{tabular}
\caption{The results of $p$--value of Wilcoxon Rank test calculated for spectral parameters for a small range of scales.}
\label{table2}
\end{table*}

\section{Conclusions}\label{s4}
This work tests the multifractal character of the sEMG signals recorded from an external anal 
sphincter at different stages of rectal cancer treatment procedure.  
For each analyzed time series, two distinct scaling regions were identified, for which 
multifractal spectra exhibit a different character.
The multifractal and monofractal nature can be seen for the regions of short and large 
time scales respectively. 
Additionally the multifractal spectra based on standard DFA were compared with EMD based one.
The later algorithm shifts the spectra towards the higher fluctuations or smaller values of the
singularity exponent. The average values of the considered spectral parameters (width and maximum)
for individual stages of treatment are respectively lower in the EMD case.
This seems to be the generic behaviour for the analysed EMG data. The changes of parameters 
between individual stages, $D_1$--$D_4$ has exactly the same 
tendency when the results of those two methods are compared.  
On top or that the source of multifractality within the short time
scales was identified as a result of the long range correlation effects for large and weak fluctuations. The statistical analysis of spectral parameters with the nonparametric Wilcoxon test distinguish state $D_3$ (6 month after the surgery).
There were statistically significant differences between $D_3$ and the rest of  stages for the spectrum width in the case of both deternding methods -- DFA and EMD.
Applied fractal methods also show the decreasing width and spectrum maximum one month after the surgical operation for all of the patients. The differences are small but the tendency 
is visible. The findings reflect also the visual variety of the raw data between the stage $D_2$ and 
all others.\\

\begin{flushleft}Conflicts of Interest: None \\
Funding: This work was partially supported by
the JUMC research grant: K/ZDS/006369.\\
Ethical Approval: Research was approved by a decision of Bioethical Commitee of Jagiellonian
University for research grant WL-ZKL-94.\end{flushleft}

\end{document}